\documentclass[acmtog, timestamp]{acmart}
\usepackage{xspace}
\usepackage{xcolor}
\usepackage{booktabs} 
\usepackage{hyperref}
\usepackage{verbatim}
\hypersetup{
    colorlinks=false,
    linkcolor=blue,
    filecolor=magenta,      
    urlcolor=cyan,
}
\setcitestyle{square}

\graphicspath{{images/}}

\usepackage{microtype} 
\usepackage{breakcites}
\usepackage{ellipsis} 
\usepackage[mathletters]{ucs}
\usepackage[utf8x]{inputenc}
\usepackage{mathtools}
\usepackage{amsbsy}
\usepackage{upgreek}
\usepackage{dblfloatfix}

\usepackage[table]{xcolor}

\usepackage{tabularx}
\usepackage{booktabs}

\usepackage{booktabs} 
\usepackage{colortbl} 
\usepackage{xcolor} 
\usepackage{xfrac}

\newcommand{\cmmnt}[1]{}

\usepackage{amsmath}

\usepackage{amsfonts}
\usepackage{amssymb}
\usepackage{enumitem}

\definecolor{lightbluishgrey}{rgb}{0.78,0.86,0.93}

\newcommand{\rinat}[1]{{{#1}}}
\newcommand{\rinatnew}[1]{{{#1}}}
\newcommand{\rinatchange}[1]{{{#1}}}

\newcommand{\library}{library\xspace} 
\newcommand{\libraries}{libraries\xspace} 

\newcommand{\piece}{piece\xspace} 
\newcommand{\pieces}{pieces\xspace} 
\newcommand{\singlepart}{part\xspace} 
\newcommand{\parts}{parts\xspace} 

\newcommand{\animation}{mesh-animation sequence\xspace} 
 
\usepackage{layouts}

\usepackage{wrapfig}

\newcommand{\refequ}[1] {Equation~(\ref{equ:#1})}

\newcommand{\reffig}[1] {Fig.~\ref{fig:#1}}

\makeatletter
\def\reffig{\@ifnextchar[{\@myreffigloc}{\@myreffignoloc}}
\def\@myreffigloc[#1]#2{Fig.~\ref{fig:#2}, \emph{#1}}
\def\@myreffignoloc#1{Fig.~\ref{fig:#1}}
\makeatother

\let\mat = \mathbf
\usepackage{xfrac}

\usepackage{amsmath}
\usepackage{amssymb}    
\usepackage{cancel}
\usepackage[T1]{fontenc}

\newcommand{\R}{\mathbb{R}}

\newcommand{\vc}[1]{\mathbf{#1}}

\newcommand{\transpose}{{\mathsf T}}

\renewcommand{\c}{\vc{c}}
\renewcommand{\d}{\vc{r}}
\newcommand{\e}{\vc{e}}

\newcommand{\p}{\vc{p}}
\newcommand{\q}{\vc{q}}

\newcommand{\w}{\vc{w}}
\newcommand{\x}{\vc{x}}
\newcommand{\y}{\vc{y}}
\newcommand{\z}{\vc{z}}
\newcommand{\A}{\mat{A}}
\newcommand{\B}{\mat{B}}

\newcommand{\D}{\mat{R}}

\renewcommand{\G}{\mat{G}}

\renewcommand{\S}{\mat{S}}

\newcommand{\W}{\mat{W}}
\newcommand{\X}{\mat{X}}
\newcommand{\Y}{\mat{Y}}
\newcommand{\Z}{\mat{Z}}

\usepackage{siunitx}
\usepackage{graphicx}

\usepackage{wrapfig}

\makeatletter
\let\save@mathaccent\mathaccent
\newcommand*\if@single[3]{%
  \setbox0\hbox{${\mathaccent"0362{#1}}^H$}%
  \setbox2\hbox{${\mathaccent"0362{\kern0pt#1}}^H$}%
  \ifdim\ht0=\ht2 #3\else #2\fi
  }
\newcommand*\rel@kern[1]{\kern#1\dimexpr\macc@kerna}
\newcommand*\widebar[1]{\@ifnextchar^{{\wide@bar{#1}{0}}}{\wide@bar{#1}{1}}}
\newcommand*\wide@bar[2]{\if@single{#1}{\wide@bar@{#1}{#2}{1}}{\wide@bar@{#1}{#2}{2}}}
\newcommand*\wide@bar@[3]{%
  \begingroup
  \def\mathaccent##1##2{%
    \let\mathaccent\save@mathaccent
    \if#32 \let\macc@nucleus\first@char \fi
    \setbox\z@\hbox{$\macc@style{\macc@nucleus}_{}$}%
    \setbox\tw@\hbox{$\macc@style{\macc@nucleus}{}_{}$}%
    \dimen@\wd\tw@
    \advance\dimen@-\wd\z@
    \divide\dimen@ 3
    \@tempdima\wd\tw@
    \advance\@tempdima-\scriptspace
    \divide\@tempdima 10
    \advance\dimen@-\@tempdima
    \ifdim\dimen@>\z@ \dimen@0pt\fi
    \rel@kern{0.6}\kern-\dimen@
    \if#31
      \overline{\rel@kern{-0.6}\kern\dimen@\macc@nucleus\rel@kern{0.4}\kern\dimen@}%
      \advance\dimen@0.4\dimexpr\macc@kerna
      \let\final@kern#2%
      \ifdim\dimen@<\z@ \let\final@kern1\fi
      \if\final@kern1 \kern-\dimen@\fi
    \else
      \overline{\rel@kern{-0.6}\kern\dimen@#1}%
    \fi
  }%
  \macc@depth\@ne
  \let\math@bgroup\@empty \let\math@egroup\macc@set@skewchar
  \mathsurround\z@ \frozen@everymath{\mathgroup\macc@group\relax}%
  \macc@set@skewchar\relax
  \let\mathaccentV\macc@nested@a
  \if#31
    \macc@nested@a\relax111{#1}%
  \else
    \def\gobble@till@marker##1\endmarker{}%
    \futurelet\first@char\gobble@till@marker#1\endmarker
    \ifcat\noexpand\first@char A\else
      \def\first@char{}%
    \fi
    \macc@nested@a\relax111{\first@char}%
  \fi
  \endgroup
}
\makeatother

\usepackage[ruled,vlined]{algorithm2e}
\usepackage{etoolbox}
\usepackage{array}

\newcommand{\figs}{}
\def\figs/{figs/}

\newcommand{\dA}{\;dA}

\DeclareMathOperator*{\argmin}{argmin}

\usepackage[verbose]{newunicodechar}
\newcommand*\Bell{\ensuremath{\boldsymbol\ell}}

\usepackage[ruled]{algorithm2e} 

\SetAlFnt{\small}
\SetAlCapFnt{\small}
\SetAlCapNameFnt{\small}
\SetAlCapHSkip{0pt}
\IncMargin{-\parindent}

\acmJournal{TOG}
\acmVolume{9}
\acmNumber{4}
\acmArticle{39}
\acmYear{2017}
\acmMonth{3}

\setcopyright{acmcopyright}

\acmDOI{0000001.0000001_2}

\acmSubmissionID{380}

\begin{document}
\title{A system for efficient 3D printed stop-motion face animation} 
\author{Rinat Abdrashitov, Alec Jacobson, Karan Singh}\affiliation{\institution{University of Toronto}}


\begin{teaserfigure}
\includegraphics[width=\textwidth]{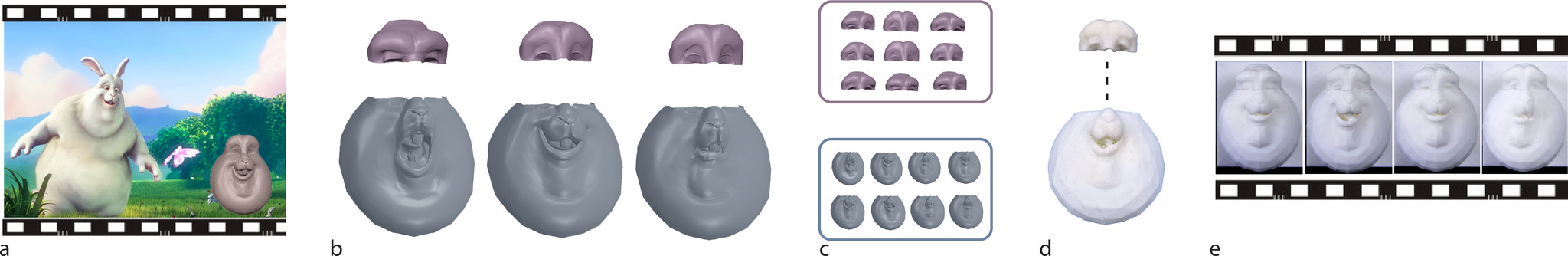}
  \caption{
    Given an input \rinatnew{\animation} (a), our system segments and deforms the 3D mesh into \parts that can be seamlessly
  joined (b). Each \singlepart is independently used to compute a replacement
  library,
  representative of the deforming \singlepart in the input (c). StopShop
  iteratively optimizes the library and a mapping from each frame of the
  input \rinatnew{animation} to the \pieces in the library. Libraries are 3D printed and
  each frame is assembled according to the optimized mapping (d). Filmed in
  sequence\rinatnew{,} the animation is realized as a stop motion film (e).}
\label{fig:teaser}
\end{teaserfigure}

\begin{abstract}
Computer animation in conjunction with 3D printing has the potential to positively impact traditional stop-motion animation. As 3D printing every frame of a computer animation is prohibitively slow and expensive, 3D printed stop-motion can only be viable if animations can be faithfully reproduced using a compact library of 3D printed and efficiently \rinatnew{assemblable} parts. 
We thus present the first system for processing computer animation sequences (typically faces) to
produce an optimal set of replacement parts for use in 3D printed stop-motion animation. 
Given an input animation sequence of topology invariant deforming meshes, our problem is to output a library of replacement parts and per-animation-frame assignment of the parts, such that we maximally approximate the input animation, while minimizing the amount of 3D printing and assembly.
Inspired by current stop-motion workflows, \rinatchange{a user manually indicates which parts of the model are preferred for segmentation; then, we find curves with minimal
deformation along which to segment the mesh}. We then present a novel algorithm to zero out deformations
along the segment boundaries, so that replacement sets for each part can be interchangeably and seamlessly assembled together. The part boundaries are designed to ease 3D printing and instrumentation
for assembly. Each part is then
independently optimized using a graph-cut technique to find a set of replacements, whose size can be user defined, or automatically computed to adhere to a printing budget or allowed deviation from
the original animation. Our evaluation is
threefold: we show results on a variety of facial animations, both \rinatnew{digital}
and 3D printed, critiqued by a professional animator; we show
the impact of various algorithmic parameters; and compare our results to
naive solutions. Our approach can reduce the printing time
and cost significantly for stop-motion animated films.

\end{abstract}
 
\maketitle
%
%
%

%
%


%

\section{Introduction}

Stop-motion is a traditional animation technique that moves a physcial object in
small increments between photographed frames, to produce the illusion of fluid
motion. As with animation in general, arguably the most expressive part of a
character is its face.  Extensive use of clay replacement \rinatnew{libraries} for dialogue and
facial expressions, goes back as far as {\it The New Gulliver 1935}. \rinatnew{The use of a replacement library has become the standard} approach to the stop-motion animation of
expressive deformable objects, \rinatnew{in particular for facial animation}. With the advent of 3D
printing, replacement animation has become a bridge between the disparate worlds
of digital computer animation and physical stop-motion, and is increasingly used
as the preferred technique for producing high-quality facial animation in stop
motion film \cite{priebe2011advanced}. 

Faces and 3D models in general are created digitally (or physically sculpted and
scanned) to produce a replacement \library that covers the expressive range of
the 3D model. This library, typically containing thousands of variations of a
deformable model is then 3D printed and cataloged. Additional post-processing
may be required, including sanding down edges, smoothing inconsistencies, and
hand painting the 3D prints. The replacement \library is then ready to be used in
stop-motion sequences \cite{alger2012art}. Alternately, the 3D model could be
entirely computer animated, and each animation frame of the model independently 3D printed and post-processed for use on a physical set.

\begin{figure}
\includegraphics[width=\columnwidth]{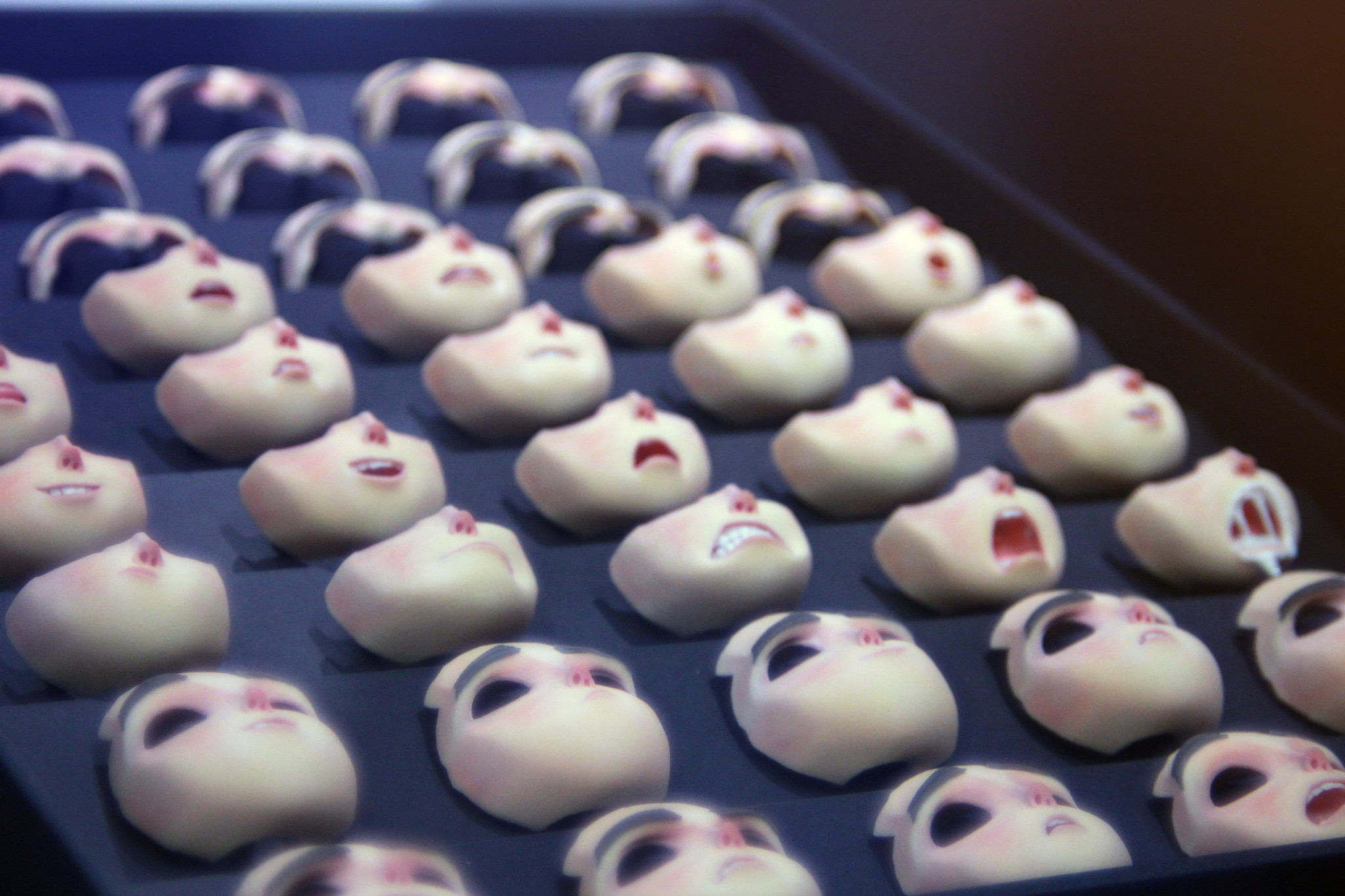}
\caption{Modern stop-motion films such as Laika's \emph{ParaNorman} amass replacement \libraries of thousands of 3D printed pieces.}
\label{paranorman}
\end{figure}

In either case, the cost in terms of printing and post-processing time,
material, storage and money is prohibitive.  Each character of Laika's
stop-motion feature film {\it Coraline} could \rinatnew{have} as many as 15,000 faces and
up to 250,000 facial expressions \cite{Kolevsohn:2009}.  {\it Paranorman}
required 8308 pounds of printer powder, and 226 gallons of ink over the course
of production \cite{priebe2011advanced} (see Figure~\ref{paranorman}). This current practice for character faces (let alone complete 3D models) is expensive for large film studios and completely beyond the reach of independent
\rinatnew{filmmakers}.

Due to the tedious nature of physically moving or replacing objects in the scene for each frame, stop motion objects are typically animated at a lower framerate (often "on twos'' or every other frame).  Some films, such as  Aardman's {\it Flushed Away} or Blue Sky's {\it The Peanuts Movie}, even opt to simulate the aesthetic appeal of stop-motion entirely, via computer animation.  As evident by these films, the slight choppiness and lower framerate can be an intentional artistic decision.
Our research addresses both 3D printing costs and animation aesthetic, providing
users with a system that can produce animation sequences in a
stop-motion style digitally, or physically with minimal 3D printing, saving printing time and material.

We present an end-to-end solution
designed to optimize the 3D printing of a replacement \library of a deformable 3D
object, such that high-quality stop-motion approximations of input computer
animations can be assembled from that \library (see
Figure~\ref{fig:teaser}). At the core of our system is an optimization problem (Section 3.3)  whose solution provides an optimal replacement \library to be 3D printed and a per-animation-frame assignment of \pieces from this \library to reconstruct the input animation faithfully.

As is common with replacement libraries \cite{priebe2011advanced}, we can amplify
the expressive range of the deformable face/object by first segmenting it into
\rinatchange{multiple parts. A user specifies the approximate location of the parts, and we calculate boundaries that have minimal or zero deformation between them}. An optimal replacement
\library is then computed independently for each part, and the object assembled by
interchangeably combining \pieces from each part's \library.
The replacement \library \pieces also need to be instrumented with connectors before 3D
printing, so that repeated object re-assembly for stop-motion, is quick and sturdy.

We propose a series of algorithms to assist in the process of creating a \library
of mix-and-matchable printable \pieces and a set of assembly instructions to
recreate a given \animation.
In particular, we introduce: a novel mesh segmentation method to find
near-stationary part boundaries, a deformation process to homogenize
part boundaries allowing temporal reshuffling of segmented parts, and
finally we simultaneously optimize for a replacement library of printable pieces and their
assignment to each frame of an input animation.

%
%

%
We evaluate our algorithm in Section 5 by showing compelling results, both
digital and 3D printed, and a comparison to a naive approach to the above problem. As shown in our accompanying video, we are able to faithully approximate input animation of cartoon characters as well as high-fidelity computer animation models. In our examples we achieve a  $\approx25x$ saving over printing each frame for animations of $\approx750$ frames.


\section{Related Work}
Our research is inspired by the challenges and animation processes at stop-motion studios like Aardman and Laika \cite{priebe2011advanced}, where 3D printing, computer modeling and animation tools are an increasingly indispensible part of the animation workflow.
Despite the popularity of stop-motion animation, the topic has received little
attention in \rinatnew{the} computer animation research literature. 
We thus focus our attention on research topics closest to our problem at an
abstract level and those similar in methodology.

\paragraph{Stop-motion}
Commercial stop-motion software such as \textsc{Stop Motion Pro} or
\textsc{DragonFrame}, \rinatnew{focuses} on optimized camera controls and
convenient interfaces for assembly and review of captured images.
There is long history of research on interfaces and techniques for performance animation, such as for  paper cut-out animations \cite{Barnes:2008:VPP}. Stop-motion  armatures have also inspired research into tangible devices \cite{Knep1995,Baecher:2016:DefSense} and interfaces \cite{wires} for posing and deforming  3D characters. Stop-motion has also been applied to study low fidelity prototyping
for user interfaces \cite{Bonanni:2009kj}. Digital removal of seams and hands from stop-motion images has been addressed by \cite{Brostow:2001:IMB}. \cite{stophands} presented a tool 
to aid in the generation of motion blur between static frames to show fast motions. 
However, the problem of generating replacement \rinatnew{libraries} for the purpose of 3D printed stop motion animation has not been addressed before.

\paragraph{Animation Compression}
Although not intentional, stop-motion can be seen as a \emph{compression} of
high-framerate or continuous animation into a much \rinatnew{smaller} set of frames.
In computer graphics and especially computer game development, there have been
many methods proposed for compressing animations: for deformable meshes using
principal component analysis \cite{AlexaM00,Sattler:2005:SEC,Vasa2014}, for
articulated characters using skeletal skinning subspaces
\cite{James:2005:SMA,Le:2014baa}, or by analyzing patterns in motion capture
data \cite{Gu:2009jg}.
These methods for digital animation are free to define interpolation operations,
effectively approximating the input animation with a continuous (albeit high
dimensional) function space. 
Stop motion in contrast requires a discrete selection: a 3D-printed face is either used for
this frame or not.
We cast this problem of extracting a printed library of shapes and assigning
those shapes to each frame of the animation as one of \emph{sparse dictionary
learning} or \emph{graph clustering}, well studied topics often used in
computer graphics.
In particular, Le \& Deng \shortcite{Le:2013kh} use sparse dictionary learning
to significantly compress mesh animations, as a weighted combination of a few basis meshes. While their weights are sparse, we must represent every animated frame using a single physical replacement mesh, \rinatnew{necessitating} a very different optimization strategy.

\paragraph{Stylization}
Much of the work in stylizing characters pertains to painterly rendering  or caricature \cite{nprstar}.
Similar to signature "choppy" style of stop-motion, controllable temporal flickering has been used to  approximate
the appearance of real hand-painted animation of faces \cite{fivser2017example} and articulated characters \cite{dvorovznak2018toonsynth}.
Video summarization techniques select discrete set of images or
clips that best sum up a longer clip, recently using deep learning to
select semantically meaningful frames \cite{otani2016video}.
Stop-motion also requires a reduced but typically larger, discrete set of replacement 3D models, not to summarize but to approximate an input animation.
%
Other research in stylizing 3D animation has explored key-pose and motion-line extraction from 3D animations for comic strip like depiction. 
Stop-motion in contrast, can be imagined as geometry "posterization" along an animation, analogous to the problem of image and video color posterization \cite{videotoon}, albeit with different objectives.
Stop-motion stylization of an animation can be also 
interpreted as the inverse problem of keyframe in-betweening
\cite{Whited:2010ec}, spacetime constraints \cite{Witkin:1988hv}, or temporal
upsampling \cite{Didyk:2010dg}. We are inspired by these methods.

\paragraph{Facial Animation}

We use replacement \library as the principal use case of replacement animation for stop-motion in this paper.
Current animation practice typically creates facial animation using \emph{blendshapes} (convex combinations of posed expressions \cite{Lewis:2014cd,Ribera2017}), with layered controls built atop to model emotion and speech \cite{Edwards:2016fm}.  The blendshape weights of a face can provide useful information regarding both the  saliency and difference in expression between faces \cite{Ribera2017}, which we exploit when available.
Our work is also inspired by work on compression using 
blendshapes \cite{Seo:2011cs} and optimization of 
\emph{spatially} sparse deformation functions \cite{Neumann:2013gb}.
In contrast, our optimization may be seen as producing an extreme form of
temporal sparsity.

\paragraph{Shape segmentation and 3D printing}

Our system also automatically segments the geometry of an
animated shape in order to maximize expressiveness of the replacement \library while
maintaining a tight 3D printing material budget. 
Shape segmentation is a fundamental and well studied problem in geometry processing \cite{Shamir:2008ui}.
In an \rinatnew{animation}, most segmentation approaches hunt
for rigid or near-rigid parts during animation 
\cite{Bergou:2007:TTD,Lee:2006cu,Ghosh:2012wv}.
Our problem is orthogonal to these; rather than looking for near-rigid \parts, 
we look for near-motionless boundaries between the segmented \parts.
Nonetheless, mesh saliency \cite{Jeong:2014fv} or other quality/printability
measures \cite{zhang:sa:2015} could easily be incorporated into our
optimization.
Segmenting and processing input 3D geometry for high quality 3D printing in general \cite{chopper, Hu:2014:APS:2661229.2661244, Herholz:2015:AFG:2816723.2816745, Wang:2016:ISQ:3058909.3058918} and faces in particular \cite{retouch} is subject to ongoing research and useful for the final 3D printing of the replacement \pieces computed by our system. \rinatchange{Instead of printing a replacement library, \citet{bickel2012physical} used material optimization methods to create synthetic silicone skin, fabricated using 3D printed molds, for animatronic figures of human faces.}

\section{System and Algorithm Design}

\begin{figure}
  \includegraphics[width=\columnwidth]{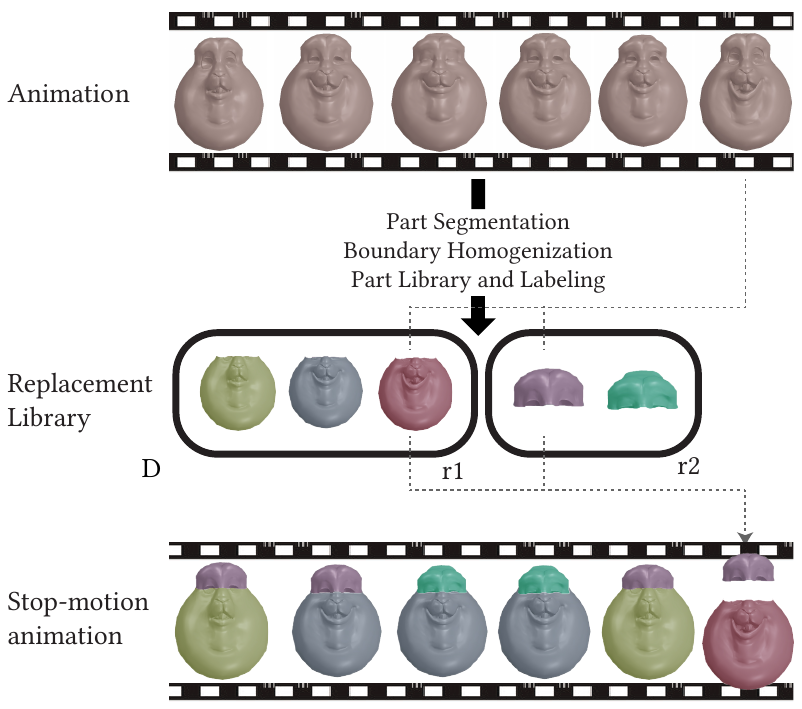}
  \caption{Algorithm overview. Input shape is segmented into parts and each animation frame is smoothly deformed so the cut has the same geometry across all frames.  For each part independently, replacement library
and corresponding assignment labels to each frame are optimized simultaneously.}
\label{fig:overview}
\end{figure}

  The input to our method is an $n$-frame \animation $\X = [\x_1 ,
  \x_2 , …, \x_n] ∈ \R^{m×3 × n}$, where $\x_f ∈ \R^{m×3}$ contains the vertex
  positions of the $f$th animation frame of a mesh with $m$ vertices and $k$
  triangles, and $\x_{fi} ∈ \R^3$ is the 3D position of the $i$th vertex in that
  frame. Multiple temporally disjoint animation clips of the mesh are simply concatenated in $\X$, with the cut locations marked. Please note that we refer to mesh faces as triangles to avoid confusion with the faces being animated, even though our solution applies to quads and other polygons.

  We assume the mesh animates via vertex displacements only and does not change
  topology, connectivity, or number of triangles ($k$) during the animation.
  The user also inputs a desired number of \parts $s$ (e.g., $s=2$ for a
  face split into top and bottom) and a desired replacement \library size $d_j \, ∀
  j= 1,\dots,s$, indicating the number of printable \pieces per \singlepart
  (e.g., $d₁=3,d₂=2$ to output 2 top face pieces and 3 bottom pieces in Figure~\ref{fig:overview}).
  %
  %

  The output of our method is $s$ replacement libraries, one for each \singlepart
 containing  the correspondingly given number of \pieces to 3D print,
  and a labeling of each of the $n$ input animations frames indicating which
  piece from each part library to place in order to recreate the frame (see Figure~\ref{fig:overview}).

  As enumerated in  Figure~\ref{fig:overview}, 
our method proceeds in three steps: 1) the input shape is segmented into $s$
  \parts{} \rinatnew{ with} a minimally noticeable cut, 2) each input frame is smoothly
  deformed so the segmentation cut across \emph{all} frames has the same
  geometry, and, finally, 3) for each \singlepart independently, \rinatnew{the} replacement \library and corresponding assignment labels to each frame are optimized simultaneously.

\subsection{Part Segmentation}
Many deformable objects like faces have localized regions of deformation
separable by near rigid boundaries, though the exact location of the cut
separating these regions is generally non-planar, curving around features like
eyes, cheeks, and noses.
Existing stop-motion facial animations often segment a head into an upper and lower face just below the eye-line, optionally with a rigid back of the head. While
our approach generalizes (via multi-label graphcut) to $s>2$, our implementation and results focus on the predominant segmentation for faces, with $s=2$.

Our input to this stage is the \animation $\X$,
and the output, a new \animation $\Y$ with $y$ triangles and a
per-triangle \singlepart assignment $\p ∈ \{1,\dots,s\}^{y}$. The output 
is \emph{geometrically equivalent} to the input, but with new vertices and
triangles ($y$ triangles instead of the input $k$ triangles) added along a smooth boundary separating the \parts.  

Users can roughly indicate desired parts by specifying a \emph{seed} triangle (or set of triangles)
$T_j$ for each \singlepart $j$. We find a per-triangle \singlepart assignment $\q_\alpha$ for each input triangle $α$ of the average mesh. The boundaries between \singlepart regions
minimize an energy that penalizes cutting along edges that move significantly during the
input animation $\X$:
\begin{align}
\label{eq:segment_eq}
  \min_{\q ∈ \{1,\dots,s\}^k} &
  ∑\limits_{α=1}^k u(α) + 
  γ ∑\limits_{α=1}^k ∑\limits_{β=1}^k b(α,β)
\end{align}
where $γ$ balances between the unary and binary terms described \rinatnew{below} (we
use a default of $γ=100$ for 3D models scaled to a unit bounding-box).
The unary \emph{data} term $u(α)$ penalizes \parts from straying in distance from the input seeds:
\begin{align}
  u(α) &:= \text{dist}(α,T_j), \text{  $j = \q_α$}
\end{align}
where $\text{dist}(α,T_j)$ measures the geodesic distance from the triangle $α$ to the closest seed in the set $T_j$. The binary \emph{smoothness} term penalizes cuts that pass
through shapes that have high displacement from their average position:
\begin{align} 
  b(α,β) &:= \begin{cases}
    ∑\limits_{f=1}^n \| \e_{fαβ} \| \left( 1 + ∑\limits_{i∈α∩β} \| \x_{fi} -
    \tilde{\x}_i\| \right) & \text{ if $q_α ≠ q_β$,} \\
    0 & \text{ otherwise,} 
\end{cases}
\end{align}
where $\tilde{\x} ∈ \R^{m×3}$ denotes the average mesh vertex positions across the animation, $\| \e_{fαβ} \|$ is the length of the edge between triangles $α$ and $β$ at frame $f$ and $α∩β$ indicates the indices 
of the shared vertices on this edge. The $1+$ penalizes long cuts even in
non-moving regions.

\begin{figure}
  \includegraphics[width=\columnwidth]{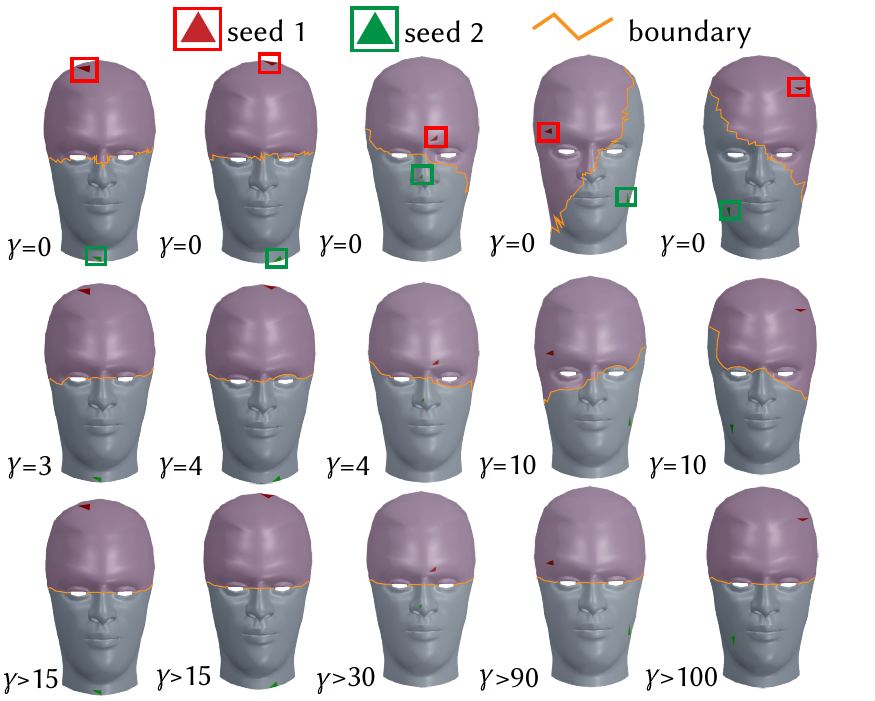}
   \caption{\rinatchange{Robustness of the part segmentation method with respect to perturbations of the seed triangle placement. Each column shows a different initialization of seed triangles with the first row showing the extreme case without the binary smoothness term as in Eq. \ref{eq:segment_eq}. Increasing the influence of the binary term via $\gamma$ produces the same boundary.}}
\label{fig:different_seed}
\end{figure}

\begin{figure}
  \includegraphics[width=\columnwidth]{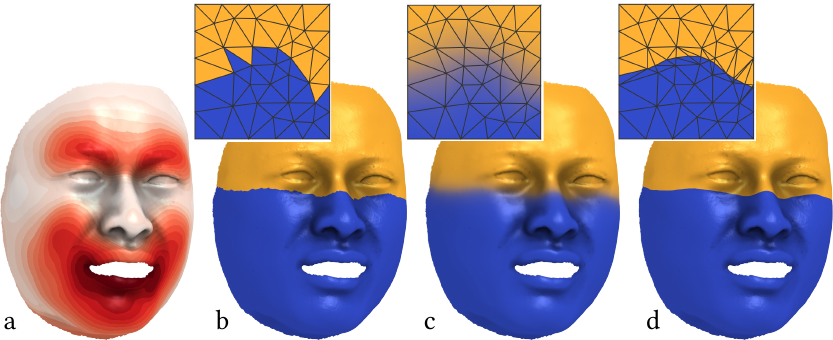}
   \caption{Left to right: average displacement visualized over the average
   face is converted into a per-triangle segmentation. We smooth this as a
   scalar per-vertex function and then extract a smoothly bounded segmentation
   as an iso-contour.}
\label{fig:bndr_smooth}
\end{figure}

This energy is efficiently minimized via graphcut-based multilabel approach \cite{Boykov2001EAE, kolmogorov2004energy, boykov2004experimental}. The result is a
per-\emph{triangle} labeling.
\rinatchange{Since the user manually chooses seed triangles by clicking on the mesh, our optimization needs to be robust to perturbations of the seed triangle placement. Figure~\ref{fig:different_seed} shows that we find the same boundary once $\gamma$ is large enough.} For a generic mesh, the \singlepart boundary may
zig-zag due to the necessity of following mesh edges (see Figure~\ref{fig:bndr_smooth}(b)). This is not only
aesthetically disappointing but pragmatically problematic: jagged boundaries
will prevent 3D printed parts from fitting well due to printer inaccuracies. 

\paragraph{\bf Part boundary smoothing}
We smooth per-triangle \singlepart boundaries by treating each \singlepart as an indicator
function ($χ_j(\x) = 1$ if triangle $\x$ is in \singlepart $j$, $χ_j(\x) = 0$ otherwise) (see Figure~\ref{fig:bndr_smooth}).
We move each indicator function into a per-vertex quantity (no longer binary) by
taking a animation-average-triangle-area-weighted average of $χ_j$ triangle values.
Treating each per-vertex quantity as interpolated values of a piecewise-linear
function defined over the mesh, we mollify each segmentation function by
Laplacian smoothing.
Because the input indicator functions partition unity, so will the output
smoothed functions: each function can be thought of as a point-wise \emph{vote}
for which \singlepart to belong to.
Finally, the smoothed \singlepart boundaries are extracted by meshing the curves
that delineate changes in the maximum vote and assigning each (possibly new)
triangle to the \singlepart with maximum value (after meshing, the maximum is
piecewise constant in each triangle).
This meshing does not change the geometry of the surface, only adds new vertices
f

%

Note that the number of vertices and triangles on the \animation $\Y$ will likely change from the number of vertices $m$ and triangles $k$ of the input \animation $\X$, as a result of the smooth part boundary extraction. In subsequent steps, for notational simplicity however, we will continue to use $m$ and $k$ to refer to the vertex and face count of the 3D meshes being processed.

\subsection{Part Boundary Homogenization}

\begin{figure}
  \includegraphics[width=\columnwidth]{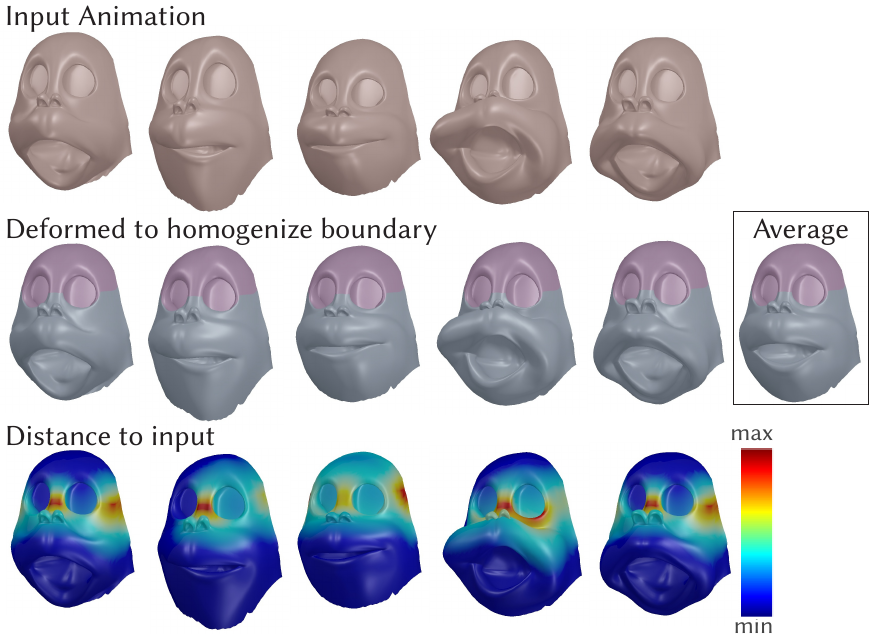}
   \caption{The input \animation is cut into two parts according to \rinatnew{the} displacement
   from the average face (inset). Each mesh is then minimally and smoothly
   deformed so \rinatnew{that} the \singlepart boundary is constant across the animation.}
\label{fig:cut}
\end{figure}

\begin{figure}
  \includegraphics[width=\columnwidth]{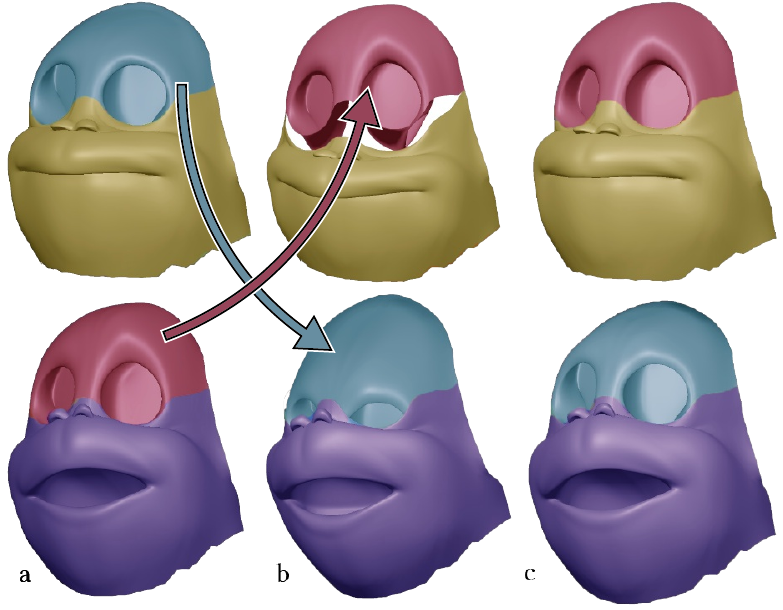}
\caption{
  In (a), two input frames (rows) are each segmented into two parts (colors).
  Simply mixing tops and bottoms of the original meshes leads to boundary
  mismatches (b).
  Instead, our system optimizes deformations for each \singlepart to
  homogenize the boundaries, allowing seamless assembly (c).}
\label{fig:random_switch}
\end{figure}

We now deform all
frames of the segmented \animation $\Y$,
so that the geometry of each \rinatnew{frame} along the \singlepart boundaries is temporally
constant (see Figure~\ref{fig:cut}).
This will allow us to mix and match different poses for each \singlepart while
maintaining continuity across the \singlepart boundaries (see Figure~\ref{fig:random_switch}).
Given a 
set of $n$ mesh positions $\Y= [\y_1 , \y_2 , …, \y_n] ∈ \R^{m×3×n}$ and a
per-triangle part labeling $\p ∈ \{1,\dots,s\}^{y}$ as input, we compute
 a vertex-deformation of these meshes with new positions $\Z =
[\z_1 , \z_2 , …, \z_n] ∈ \R^{m × 3 × n}$.

We find a minimal deformation of the input frames by solving a constrained
optimization problem so that the \emph{displacements} move each vertex
$\y_{fi}∈\R^3$ along the
part boundaries (i.e., vertices incident on triangles with different
assignment) to its average value across the input meshes $\tilde{\y}_{i} ∈
\R^3$ and move non-boundary vertices smoothly.
We conduct this optimization for each input mesh of the sequence $\Y$.
In the continuous setting, we model this as a minimization of the
squared-Laplacian energy of the displacement field:
\begin{align}
\label{eq:def_energy}
  \argmin_{\z_f}      & ∫_{\text{surface}} \|∆(\z_f-\y_f)\|_F^2 \dA \\ 
  \text{subject to: } & \z_f = \tilde{\y}_f \text{ along the seam} \\
  \label{eq:def_energy_grad}
  \text{  and }       & ∇\z_f = ∇\tilde{\y} \text{ along the seam},
\end{align}
where the gradient condition not only ensures a unique solution, but also forces
the normal of the resulting meshes to vary consistently. This condition is of
practical importance for final fabrication: each part can be extruded inward
along its normal direction to create a boundary-matching volumetric (printable)
shell.

In practice, we implement this for our discrete triangle meshes using the
mixed Finite-Element method \cite{Jacobson:MixedFEM:2010} (i.e., squared
cotangent Laplacian). We implement the gradient condition by fixing one-ring of
vertex neighbors along the seams to their average values as well.

\rinat{The Laplacian energy (\ref{eq:def_energy}) is discretized using linear FEM Laplacian $M^{-1}L$ where $M$ is the mass matrix and $L$ is the symmetric cotangent Laplacian of the average mesh.
\begin{align}
  ∫_{\text{surface}} ||∆(\z_f-\y_f)||_F^2 \dA  & = tr((M^{-1}LD)^TMM^{-1}LD) \\ 
  & = tr(D^T(L^TM^{-1}L)D)  
\label{eq:discrete_deform}  
\end{align}
\begin{align}
\text{where } D_{f} := \begin{cases}
     \tilde{\y}_f - \y_f & \text{along seam/one-ring neighbourhood} \\
     \z_f - \y_f  & \text{otherwise}
\end{cases}
\end{align}}

\rinat{The energy term (\ref{eq:discrete_deform}) is quadratic in the unkwons $\z_f$ and convex with linear equality constraints that is solved using \textsc{Eigen's} \cite{eigen} sparse Cholesky solver.}

Though each frame's deformation is computed independently, we have modeled this
as a smooth process and, thus, the temporal smoothness of the input meshes will
be maintained: temporally smooth input animations remain smooth.

\subsection{Replacement Library and Per-Frame Assignment}
Sections 3.1 and 3.2 allow us to maximize the expressivity of a replacement \library by segmenting and deforming the input mesh into $s$ parts, whose individual replacement libraries can be arbitrarily assembled together.  Replacement \libraries for each of the $s$ parts can thus be computed independently. We now focus on 
determining the \pieces that compose the replacement \library of each \singlepart, and a per-animation-frame assignment of \pieces from these \libraries to reconstruct the input \animation faithfully.

For brevity of notation, we denote the input to this subroutine as desired \library
size $d$ and a (sub-)mesh animation of a single \singlepart $\Z = [\z₁,\z₂,…,\z_n] ∈
\R^{3m × n}$. We will operate on $\Z$ as a 2D matrix (we stack $x$,$y$, and $z$
coordinates vertically).
Optionally, the user may provide a vector $\w ∈ \R^m$ of saliency weights, so
that $w_i ∈ \R$ contains a larger (smaller) value if the $i$th vertex is more
(less) salient.
Saliency can be animator-defined, or computed automatically from criteria,
such as frames that are blendshape extremes, motion extrema
\cite{Coleman:2008}, or viseme shapes \cite{Edwards:2016fm}.
Additionally, as already mentioned, the user may optionally include a ``cut'' vector $\c ∈
\{\text{true},\text{false}\}^n$ indicating whether each frame is beginning a
new unrelated sequence in the animation (e.g., a scene change).

The output is a replacement \library of $d$ \pieces for the \singlepart $\D =[\d_1, \d_2,..,\d_d]∈\R^{3m × d}$ and a sequence of $n$ labels assigning each input frame to a corresponding \piece from the \library $\Bell = [\ell_1 , \ell_2 , … , \ell_n] ∈ \{1,…,d\}^n$. We optimize for $\D$ and
$\Bell$ to best approximate the input geometry \emph{and} the change in input
(e.g., the discrete velocity) between consecutive frames for inputs that come
from animation clips.

Our optimization searches over the continuous space of \library \pieces and the discrete space of assignment labels, to optimize the combined
$L²$ geometry and velocity energy function $E$:
\begin{align}
  \label{equ:minimization}
  \mathop{\text{minimize}}_{\D,\Bell} \ 
  \underbrace{
  ½||\X -  \D \S(\Bell)||_\W^2 + 
  \frac{λ}{2}||\X\G - \D \S(\Bell) \G||_\W^2}_{E(\D,\Bell)}, \\
  \text{where } \S(\Bell)∈\{0,1\}^{d×n},\S_{kf}(\Bell) := \begin{cases}
     1 & \text{if } \ell_f = k, \\
     0 & \text{otherwise,}
  \end{cases}
\end{align}
where $\W ∈ \R^{3n×3n}$ is a matrix containing the per-vertex saliency weights
$\w$ repeated along the diagonal for each spatial coordinate and 
$λ$ balances between shape accuracy and velocity accuracy, and $\G ∈
\R^{n×(n-1)}$ is a sparse matrix computing the temporal forward finite
difference:
\begin{equation}
  \label{equ:define-G}
  G_{fg} = \begin{cases}
    \hphantom{-}0 & \text{if } \c_f \text{ or } \c_g, \\
    \hphantom{-}1 & \text{else if } f = g+1, \\
             {-}1 & \text{else if } f = g,   \\
    \hphantom{-}0 & \text{otherwise.} 
  \end{cases}
\end{equation}

As opposed to \emph{soft} labeling \cite{wright2010sparse, elad2006image}, our labeling is \emph{hard} in the sense
that the implied stochastic ``representation'' matrix $\S(\Bell)$ is \emph{binary}. 
We are \emph{literally} going to print our replacement libraries.
This is considerably harder to optimize than the standard sparse dictionary
learning problem where sparsity is enforced via an objective term and may be
convexified using an $L^1$-norm.
Instead, we optimize using block coordinate descent.
We repeatedly iterate between:
\begin{itemize}
    \item finding the optimal replacement \library pieces $\D$ holding the
labels $\Bell$ fixed, and
    \item finding the optimal labels holding the library
fixed.
\end{itemize}

Since fixing the labels $\Bell$ also fixes the representation matrix $\S(\Bell)$,
finding the optimal \library amounts to minimizing a quadratic least squares
energy. The optimal \library $\D$ is a solution to a large, sparse linear
system of equations:

\begin{align}
  \underbrace{(\S(\Bell)\S(\Bell)^\transpose +λ\S(\Bell) \G \G^\transpose \S(\Bell)^\transpose)}_\A \D^\transpose =
  \underbrace{(\S(\Bell)+λ\S(\Bell)\G)\X^\transpose}_\B,
\end{align}
Where $\A ∈ \R^{n×n}$ is a sparse matrix and $\B ∈ \R^{n×3m}$ is a dense matrix
whose columns correspond to specific vertex coordinates.
This formula reveals that each vertex-coordinate (column in $\D^\transpose$) is
computed independently, hence, the saliency weights $\W$ fall out during
differentiation.

As long as $\S(\Bell)$ contains at least one non-zero entry per-row (i.e., each
library instance is used at least once),  the system matrix $\A$ can be
efficiently factorized (e.g., via Cholesky with reordering) and then applied
(e.g., in parallel) to each column of the right-hand side.

Fixing the library $\D$ and optimizing for the labels $\Bell$ is more
complicated, but nonetheless well posed.
%
We may rewrite the objective function $E$ in
\refequ{minimization} as a sum of \emph{unary} terms involving the independent
effect of each label $\Bell_f$ and \emph{binary} terms involving the effect of
pairs of labels $\Bell_f$ and $\Bell_g$ corresponding to the $f$th and $g$th
animation frames: 

\begin{align}
  E(\D,\Bell) &=
  ∑\limits_{f=1}^n \overline{u}(f) +
  ∑\limits_{f=1}^n∑\limits_{g=1}^n \overline{b}(f,g),\\
  \intertext{where}
  \overline{u}(f) &:= ½||\x_f  - \d_{\ell_f} ||_\W^2, \label{eq:unary_t} \\
  \overline{b}(f,g) &:= 
  \begin{cases*} 
      \frac{λ}{2} ||(\x_g - \x_f) - (\d_{\ell_g} - \d_{\ell_f})||_\W^2 & \text{if $|f-g|=1$}, \label{eq:binary_t}\\ 
      0 & \text{otherwise.} 
  \end{cases*} 
\end{align}

The binary term $\overline{b}(f,g)$ satisfies the \emph{regularity} requirement described
by Kolmogorov and Zabin
\shortcite{kolmogorov2004energy}. Specifically in the case of neighboring
animation frames with $|f-g|=1$, the term sastisfies:
\begin{align}
  \footnotesize
  \begin{array}{c}
    ½\|(\x_g - \x_f) - (\d_{\ell_f} - \d_{\ell_f})\|_F^2 \ + \\
    ½\|(\x_g - \x_f) - (\d_{\ell_g} - \d_{\ell_g})\|_F^2
  \end{array}
  &≤
  \footnotesize
  \begin{array}{c}
    ½\|(\x_g - \x_f) - (\d_{\ell_f} - \d_{\ell_g})\|_F^2 \ + \\
    ½\|(\x_g - \x_f) - (\d_{\ell_g} - \d_{\ell_f})\|_F^2 \\
  \end{array}
\end{align}
which after simplification is equal to 
\begin{align}
  \label{equ:g_cut_cond}
  0 &≤ \|\d_{\ell_g} - \d_{\ell_f}\|_F^2.
\end{align}
Since Equation (\ref{equ:g_cut_cond}) is always true we satisfy the regularity
requirement for energy to be graph-representable.
Problems of this form are efficiently solved using graphcut-based multilabel
optimization (e.g., $α$-expansion) \cite{Boykov2001EAE, kolmogorov2004energy, boykov2004experimental}. 

\begin{figure}
    \includegraphics[width=\columnwidth]{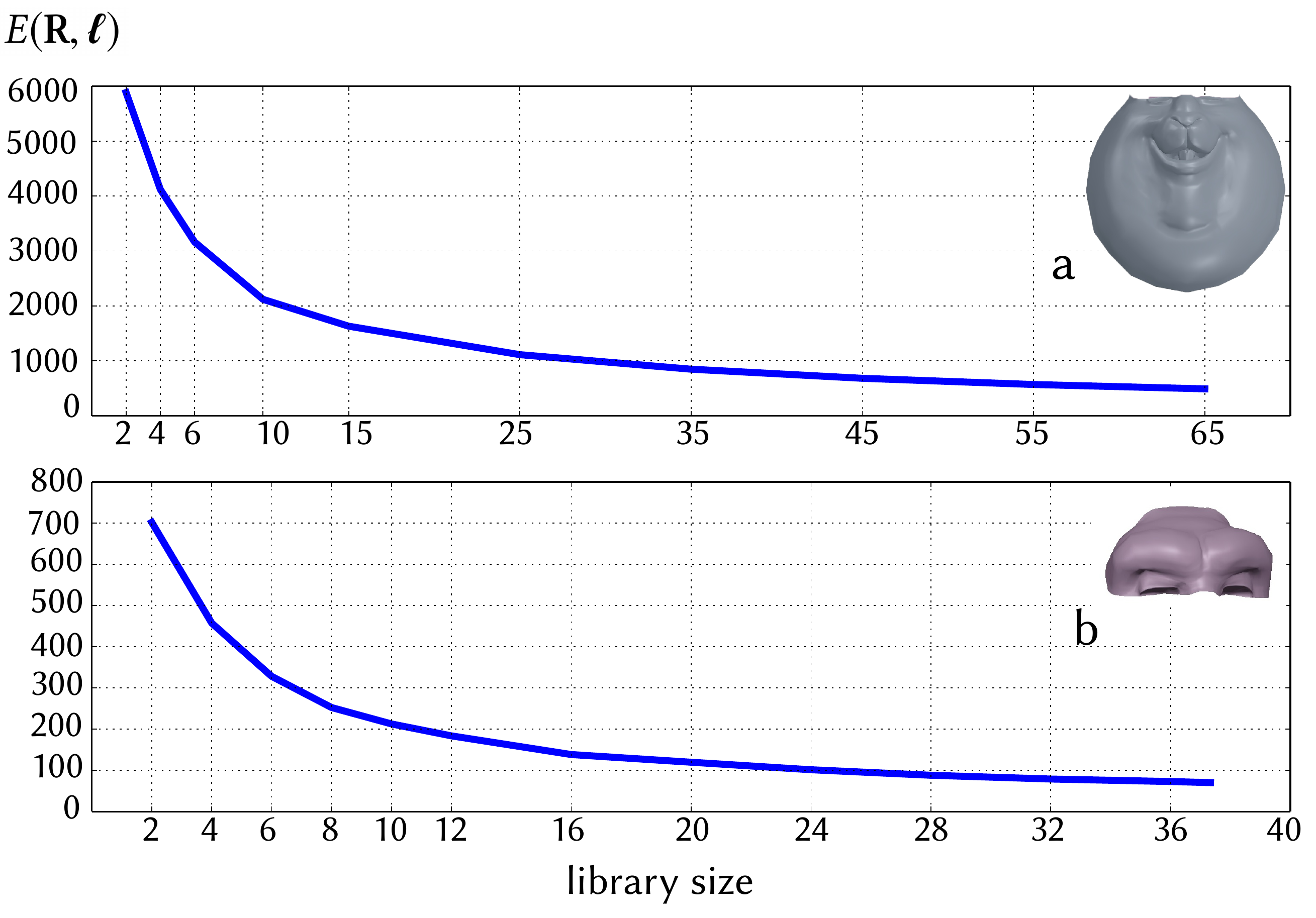}
    \caption{Increasing the number of pieces improves accuracy of the approximation but increases the cost of printing. }
    \label{fig:energy_per_label}
\end{figure}

\begin{figure}
    \includegraphics[width=\columnwidth]{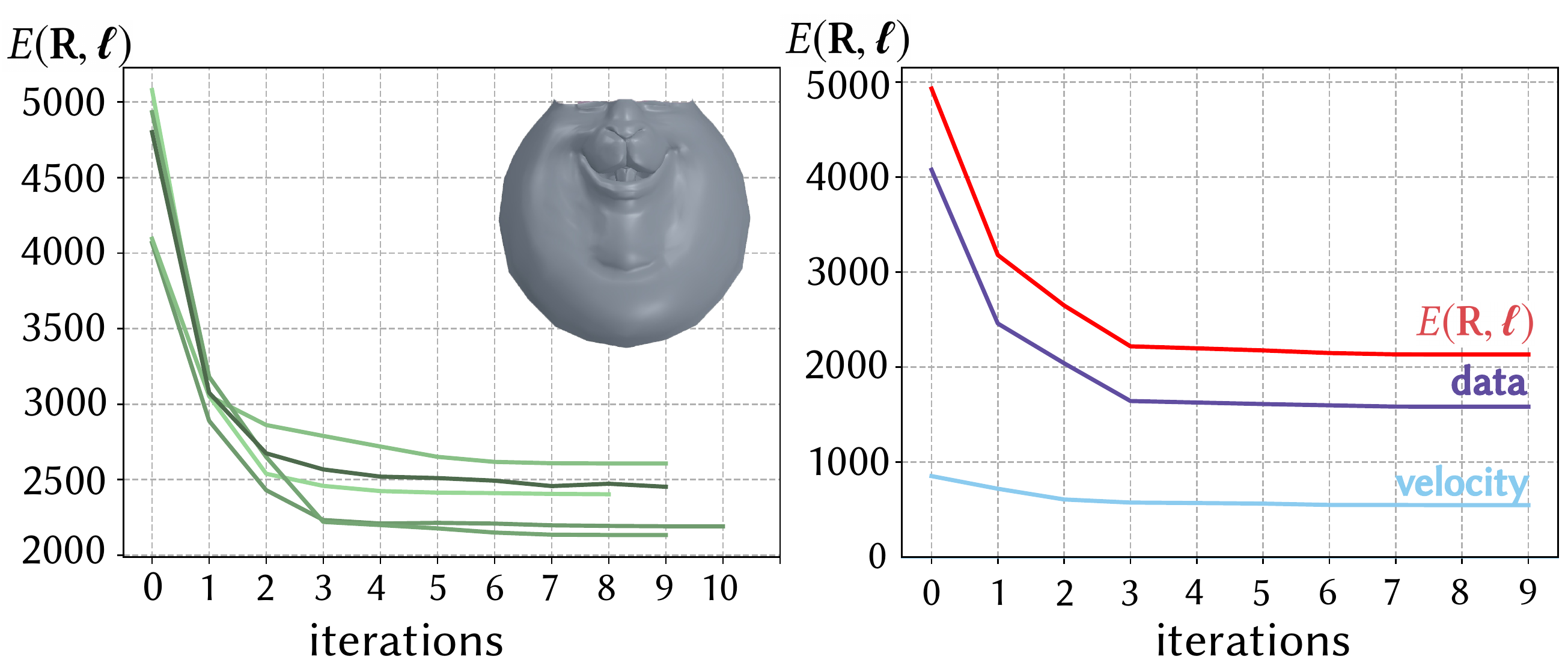}
    \caption{\rinatchange{Left: the total labeling energy $E(\D,\Bell)$ (integrated over 766 frames and 5826 vertices) for multiple runs of the block coordinate descent algorithm on a bunny sequence, 25 labels and $λ=2$ to approximate the lower part of the bunny. On average, the difference between a vertex on the input frame and its corresponding vertex on the library \piece is 0.11mm, and its per-frame velocity difference is 0.04mm. The bunny model was scaled to be $2 \times 2.5 \times 1.42$ cm, similar to printing results shown in the accompanying video. Right: single iteration of the block coordinate descent algorithm showing the data and velocity terms separately.}}
    \label{fig:bcd}
\end{figure}

When we set the velocity term weight to zero ($λ=0$), graphcut becomes
unnecessary: optimizing labels becomes a simple closest point problem, and
optimizing for the \library becomes a simple center of mass computation.
Without the velocity term, our block coordinate
descent thus, reduces to Lloyd's method for solving the $k$-means clustering problem \cite{Lloyd82leastsquares}.
In other words, for $λ>0$ we solve a generalization of the $k$-means clustering
problem, and like $k$-means, our objective landscape is
non-convex with many local minima.  Our optimization deterministically finds a
local \rinatnew{minimum} given an intial guess. We thus run multiple instances of our algorithm, with random initial assignments and keep the best solution \rinatnew{Figure~\ref{fig:bcd}.}

We now discuss practical workflow scenarios and how they fit into the above
algorithm.

\paragraph{Pre-defined replacement \library}
Sometimes the entire \library or part of it may be fixed, for example if it was
previously printed for a film prequel. Our algorithm can trivially be
used for labeling a fixed library to input animation, and a partially
specified library simply constrains the pre-defined replacements in the
library. Animators can also pick an appropriate library size $d$ based on a
visualization of the library size versus representational error (Eq. ~\ref{equ:minimization}) (see Figure~\ref{fig:energy_per_label}).

\begin{figure}
    \includegraphics[width=\columnwidth]{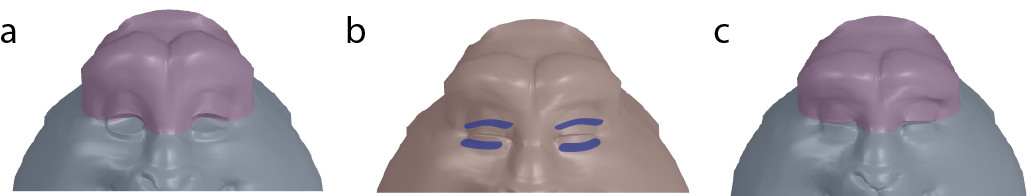}
    \caption{ Without saliency weights optimization sometimes fails to produce stop motion results that  close the eyes (a). Saliency weights are increased around eyelids (b) to have eyes properly close (c).}
  \label{fig:weights_perception}
\end{figure}

\paragraph{Arbitrary mesh animations}
Our algorithm is agnostic to the shape representation of the object $O$ in $\X$, as long as we can compute similarity functions of shape and velocity on the shape parameters.
By nature of the artistic construction of blendshapes, the $L₂$ norm of the difference of blendshapes approximates a perceptually meaningful metric. $L₂$ vertex position error in contrast may need to be augmented by  area-weighting and/or 
per-vertex rescaling according to a user painted importance function or
automatically computed mesh saliency \cite{Jeong:2014fv}.
\paragraph{Saliency weights}
Saliency \rinat{weights guide optimization to better approximate small but perceptually important regions of deformation. \rinatnew{The} amount of deformation that happens in the mouth region(lips, inner mouth and tongue) ends up taking priority over important regions like eyelids which results in lack of blinking. Figure~\ref{fig:weights_perception} illlustrates how users can manually paint saliency weights (similar to skinning weights for articulated characters) in order to ensure eyelid movement is well aproximated in the stop motion library.}

\begin{figure*}
    \includegraphics[width=\textwidth]{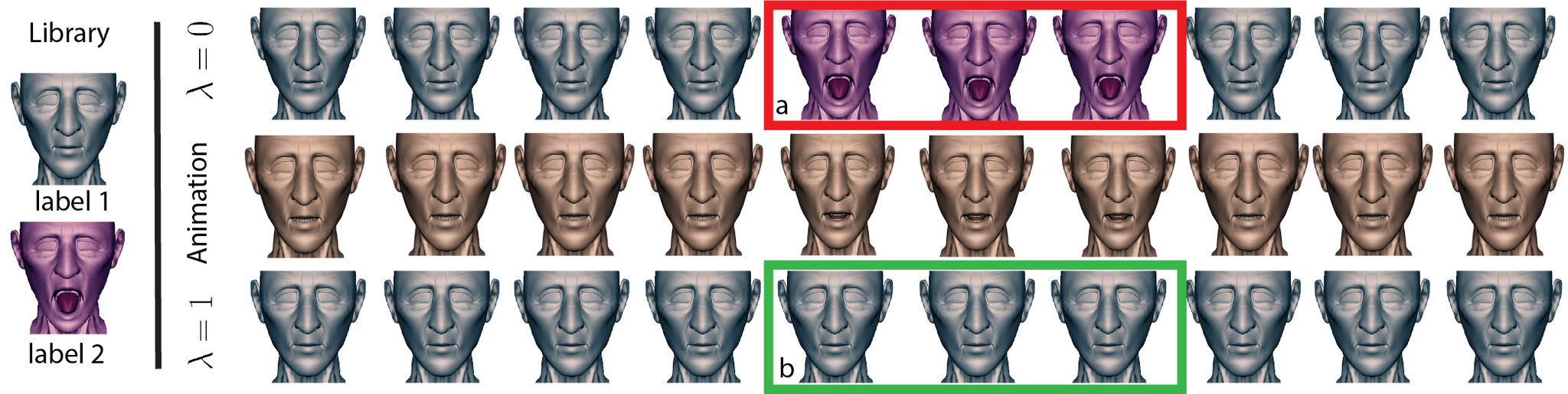}
    \caption{\rinat{A mouth opening animation (middle) is
    approximated using 2 pieces. Without the velocity term (top) the few frames (a) where the character slowly opens his mouth are labeled with the 
    extreme open mouth, resulting in a popping artifact. Including the  velocity term (bottom) prevents the popping (b).}}
    \label{fig:velocity_smile}
\end{figure*}

\begin{figure*}
    \includegraphics[width=\textwidth]{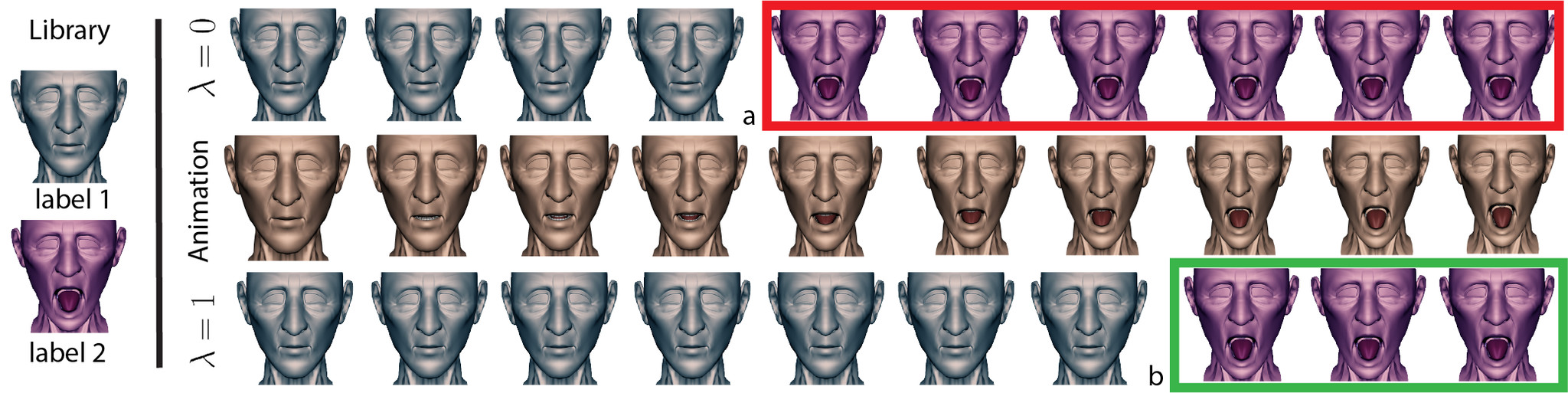}
    \caption{\rinat{An open-mouthed surprise animation (middle) is
    approximated using 2 \pieces. The replacement and labeling without velocity term (top) snaps the face to an open mouthed surprise too early (a). Including the velocity term (bottom) better preserves the emotional timing (b) of the original animation.}}
    \label{fig:open_mouth_vel}
\end{figure*}

\paragraph{Object Velocity}
The velocity term (see Equation 3, 4) is critical in preserving the smoothnes and correct timing of transitions between the object in \rinatnew{the input}. This is especially evident when \rinatnew{the} \library size is much smaller than the number of frames being approximated. Absence of this term can result in both spatial popping (see Figure~\ref{fig:velocity_smile}) and temporal sliding (see Figure~\ref{fig:open_mouth_vel}).

\rinat{Figure~\ref{fig:velocity_smile} illustrates a character 
gradually opening his mouth. Given a replacement \library of two \pieces 
({\it closed} and {\it open} mouth), our approach would correctly label the animation, 
while without the velocity term, we may see short glitches, where the label snaps to an 
{\it open} mouth creating an undesired popping effect.}

\rinat{Figure ~\ref{fig:open_mouth_vel} shows a sudden open mouthed moment of surprise animation. Without the velocity term, both the emotional onset and spatial apex of the input \animation is lost, i.e. the mouth opens earlier than it should and wider, whereas this is preserved with our approach}.


\subsection{Part Assembly}

\begin{figure}
    \includegraphics[width=\columnwidth]{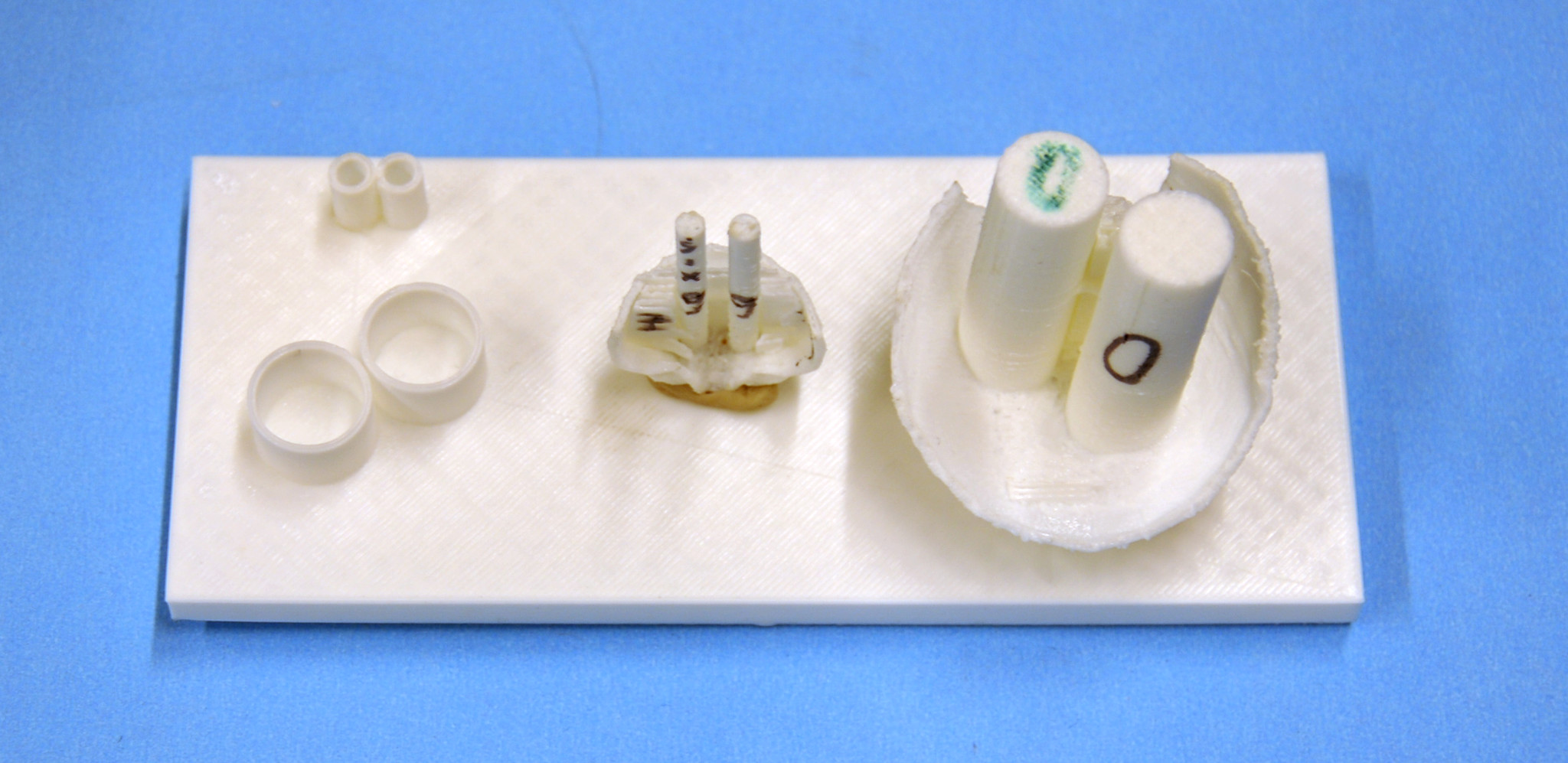}
    \caption{We create male/female plugs to connect the 3D printed ``jigs''.}
    \label{fig:jigs}
\end{figure}

Our part segmentation algorithm in Section 3.1 does not guarantee that the
object can be physically re-assembled \cite{chopper} and we do not implement any way of holding parts together. \rinatchange{Fortunately, in our experiments, the segmentation step has always produced parts that could be assembled after printing. Along the boundaries, the assemblability is locally guaranteed since the gradient condition in Eq. \ref{eq:def_energy_grad} ensures that the normal along the segmentation boundary varies consistently. Global assemblability (see, e.g., \cite{song2012recursive}), though not an issue for our examples, could be an interesting avenue for future research.
}
\rinatnew{
Most studios design custom rigs to hold stop motion parts together in order to ensure that they can be quickly and sturdily swapped out on set. For example, Laika uses magnets slotted into parts which enable animators to quickly swap different parts during the filming process.
Rather than assume a particular rig type, we did not focus on the generation of connectors between parts.
To realize our experiments, we simply created male/female plugs on parts that connect; these plugs can be fused with the part and 3D printed  (see Figure~\ref{fig:jigs}).}

where the gradient condition not only ensures a unique solution, but also forces
the normal of the resulting meshes to vary consistently. This condition is of
practical importance for final fabrication: each part can be extruded inward
along its normal direction to create a boundary-matching volumetric (printable)
shell.

In practice, we implement this for our discrete triangle meshes using the
mixed Finite-Element method \cite{Jacobson:MixedFEM:2010} (i.e., squared
cotangent Laplacian). We implement the gradient condition by fixing one-ring of
vertex neighbors along the seams to their average values as well.

\section{Implementation }
Our system has been implemented as an integrated standalone application.
The algorithms described in Section 3 were implemented in C++ using
\textsc{Eigen} \cite{eigen} and \textsc{libigl} \cite{libigl}.
Our optimization relies on a random seed\rinatnew{. We} run multiple instances, choosing
the best. Each instance takes roughly \rinatnew{5-10} iterations. The entire optimization
usually takes around 15-20 seconds for short clips of up $1000$ frames (Table ~\ref{table:stats}). Perfomance was measured on a computer with Interl Xeon CPU @ 2.40GHZ, Nvidia GTX1080 and 64GB of RAM.

\begin{figure}
    \includegraphics[width=0.48\columnwidth]{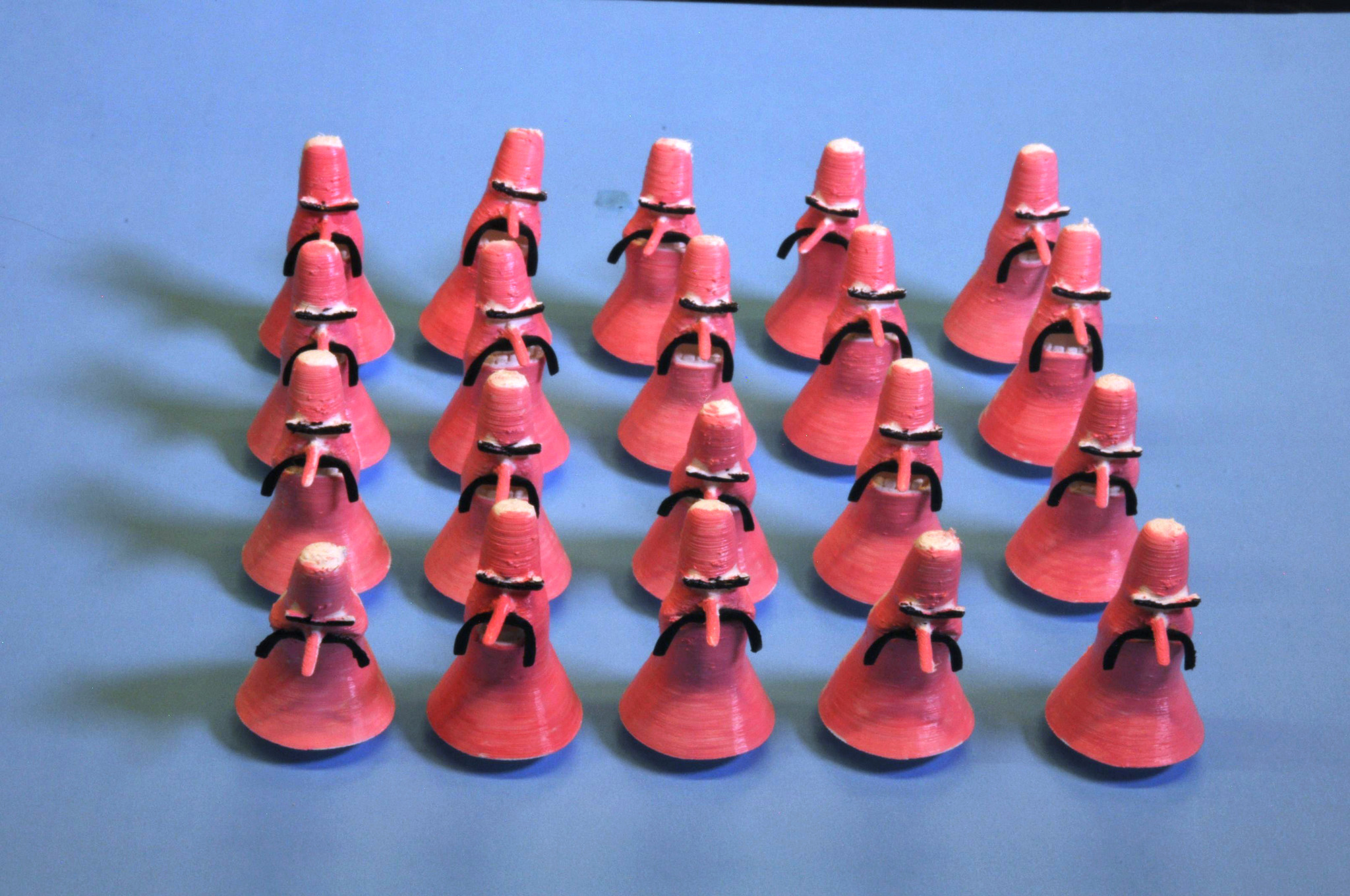}
    \includegraphics[width=0.50\columnwidth]{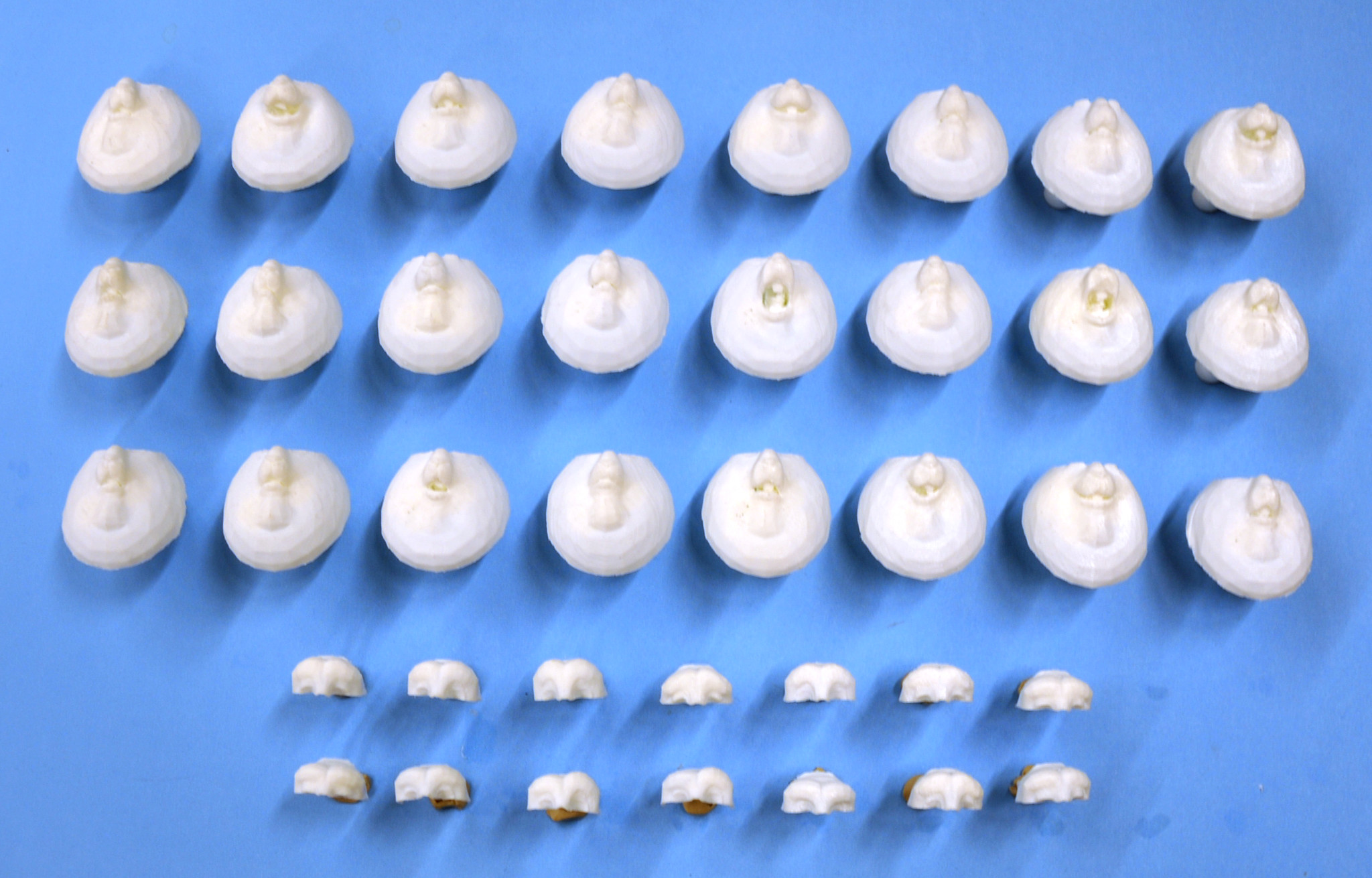}
    \caption{
      For our physically realized results, we generated 20 cartoon characters
      and 25 lower bunny faces and 15 upper halves.}
    \label{fig:bloby_library}
  \end{figure}

The digital library of parts generated using our method was 3D printed with the
DREMEL 3D20 3D printer using white PLA material. We manually colored some parts
with water based markers. Using a color powder 3D printer will certainly improve the
rendered appearance of our results (see Figure~\ref{fig:bloby_library}). 

\begin{figure}
\includegraphics[width=\columnwidth]{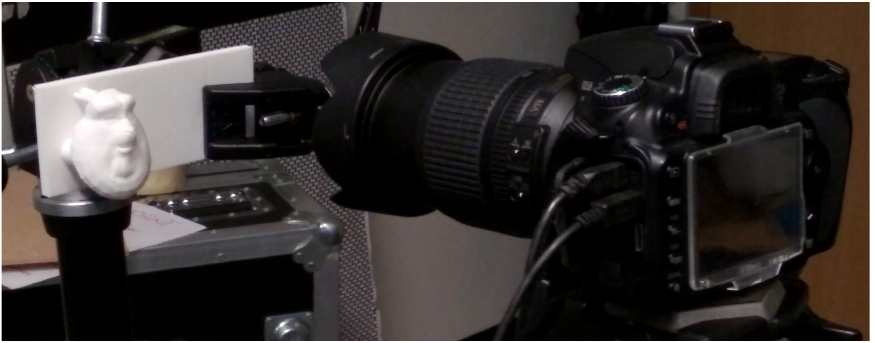}
\caption{
  We recreate a typical low-budget stop motion camera setup. }
\label{fig:camera_setup}
\end{figure}

Each printed \piece is assigned a unique ID and for every frame of the input
animations we assign the part IDs mapped by our method in Section 3. The \pieces
are connected using simple connectors or magnets (Figure~\ref{fig:jigs}). We use
a Nikon D90 DLSR camera that is controlled by Stop Motion Pro Eclipse software,
to view the scene, capture still images and mix stored frames with the live view
(see Figure~\ref{fig:camera_setup}).  Maintaining precise lighting across
sequentially shot frames can be challenging in a research lab and this is
evident in our stop-motion clips in the accompanying video.

\section{Results and Discussion}

\begin{figure}[h]
  \includegraphics[width=\columnwidth]{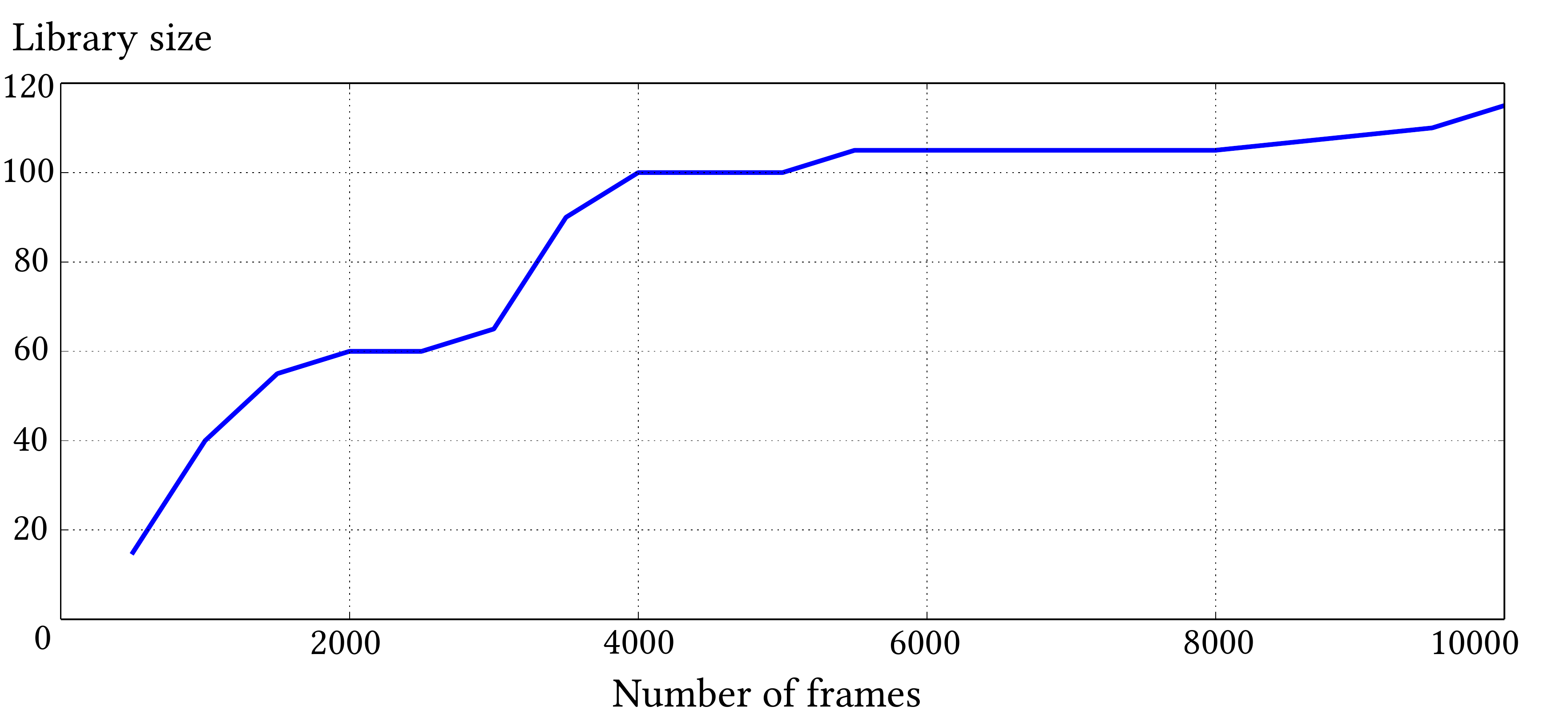}
  \caption{Number of \pieces needed to be 3D printed in order to achieve a maximum error threshold per frame.}
  \label{fig:num_of_heads_to_reach_error}
\end{figure}

\begin{figure}
  \includegraphics[width=\columnwidth]{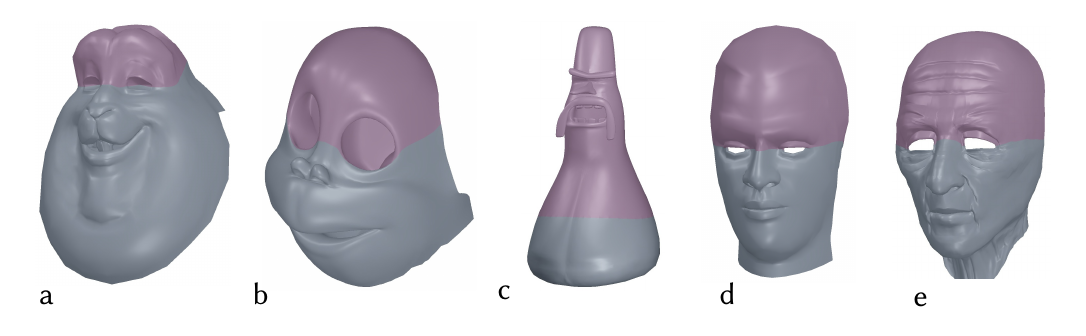}
  \caption{Our method works on cartoon characters (a, b, c) as well as high-fidelity computer animation models (d, e).}
  \label{fig:faces}
\end{figure}

\begin{figure}
  \includegraphics[width=\columnwidth]{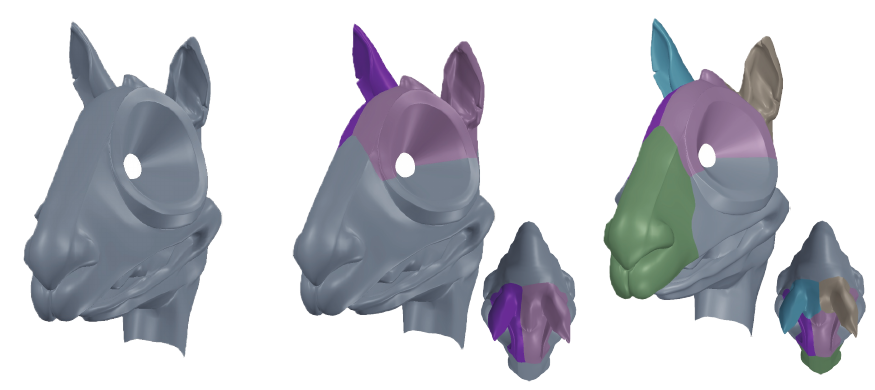}
  \caption{\rinatchange{We demonstrate the generality of our system with 3 (middle) and 6 (right) parts segmentation of the input model (left).}}
  \label{fig:faces_multi}
\end{figure}

\rinatchange{Figures~\ref{fig:faces} and \ref{fig:faces_multi}} show a variety of faces we animated using our approach (see accompanying video). 
Even for a single short animation clip, we are able to
capture $\approx 750$ frames using two replacement \libraries (20+30 pieces), a
$25x$  saving over printing each frame. 
\rinatchange{Increasing the number of parts allows to achieve comparable results while decreasing the number of pieces per part. Smaller libraries require less material leading to large cost savings and shorter printing times. For example, given the same budget of 25 pieces per part, we are able to achieve better results with the 6 parts segmentation than the 3 part segmentation or no segmentation at all (Figure ~\ref{fig:faces_multi}).}

We informally evaluated our approach with a
professional animator, who liked the ability to explore the trade-off between
animation quality and replacement library size, and felt the method captured the emotional range of the characters well, even for small \libraries.


%

\begin{figure*}
  \includegraphics[width=\linewidth]{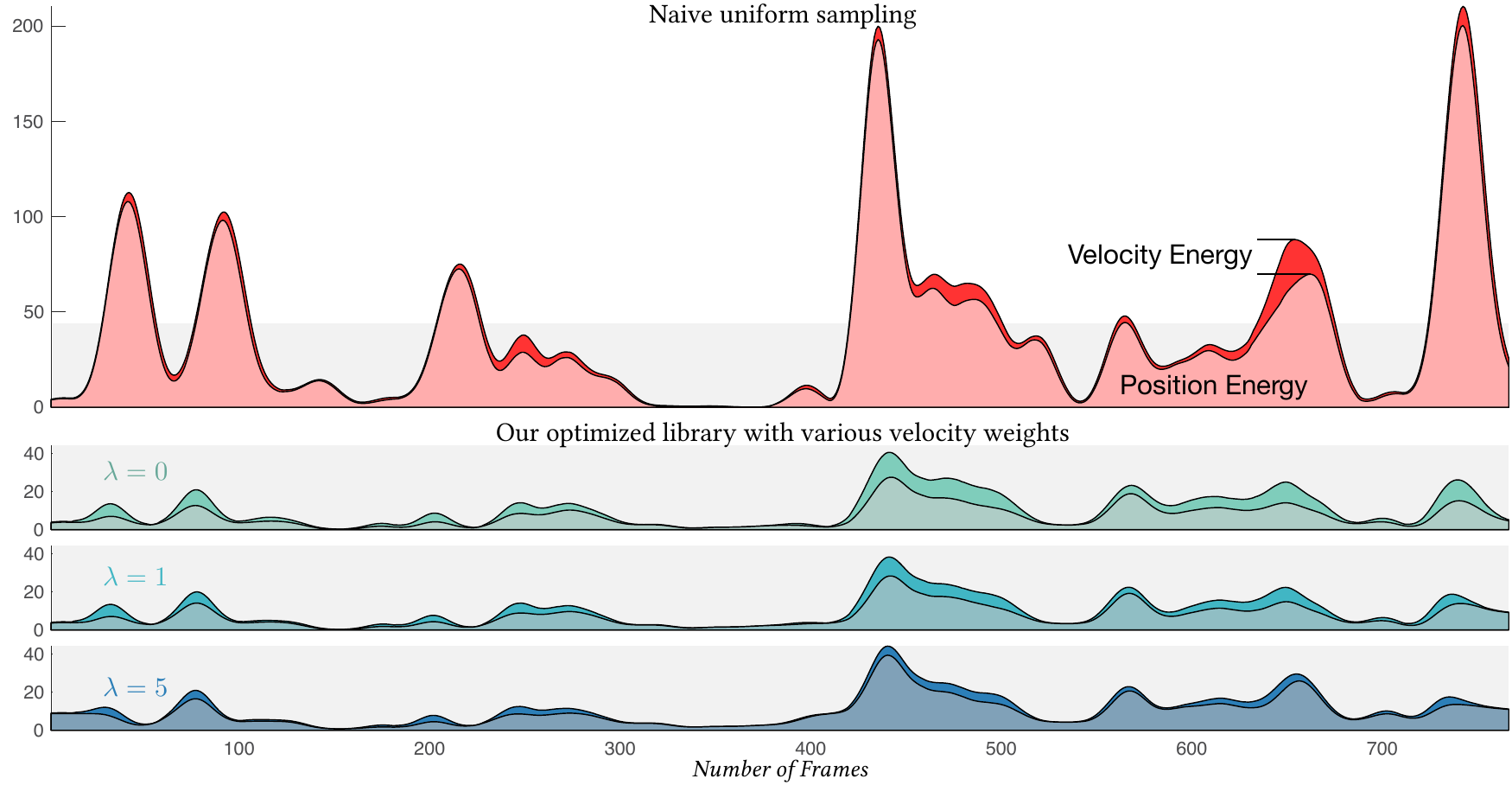}
  \caption{Top: for comparison, we fix the \library to a uniform
  sampling of the deforming object over time (and then use our assignment optimization). Uniform sampling
  produces very high position error (desaturated area) and high velocity error
  (saturated area). For visualization, we smooth temporally with a small
  windowed Gaussian.
  Below: our optimized \library produces dramatically lower error (vertical axis
  cropped to small gray box). Increasing the velocity weight $λ$ intuitively
  \emph{trades} geometric velocity error for position error.  }
  \label{fig:data_error}
\end{figure*}

\begin{table}[h!]
 \begin{center}
 \begin{tabular}{lllll} 
 \toprule 
 Model & Verticies & Frames & Labels & Time \\ [0.5ex] 
 \midrule
 \rowcolor{black!20} Monkey & 9585 & 2653 & 150 & 39 \\ 
  Bunny & 11595 & 5177 & 200 & 152 \\
 \rowcolor{black!20} Oldman & 21484 & 260 & 20  & 1   \\
 Blobby & 60464 & 229 & 20 & 4    \\[1ex]
 \bottomrule
\end{tabular}
\end{center}
\caption{Perfomance statistics. Time (in seconds) includes labeling and update steps times described in \rinatnew{Section} 3.3.}
\label{table:stats}
\vspace{-2em}
\end{table}
We compare our replacement \pieces selection and mapping algorithm in Section 3.2 to naive uniform downsampling. Quantitatively, for a 750 frame animation captured using 20 replacement \pieces, the overall error for both uniform sampling and our velocity independent ($\lambda=0$) approach is significantly higher than the velocity aware ($\lambda>0$) approach (see Figure~\ref{fig:data_error}). While the error in object shape in Figure~\ref{fig:data_error}a is comparable or marginally worse for our velocity aware over velocity independent approach, as illustrated in Section 3.2, the velocity error for $\lambda=0$ case in Figure~\ref{fig:data_error}b is understandably large.
Qualitatively, Figures~\ref{fig:velocity_smile}, \ref{fig:open_mouth_vel} show the velocity term to be critical to both
the selection and mapping to replacement pieces. 

Printing time and cost can be prohibitive if the size of the \library increases linearly with the number of frames of animation \cite{priebe2011advanced}.  
In Figure ~\ref{fig:num_of_heads_to_reach_error}, we calculate number of replacement \pieces needed in order to stay below a certain per frame error threshold, for 10,000 frames of an animated face reading chapters from the book Moby Dick.  
Given the labeling and a number of frames, we compute the minimum error value of the sum of the unary and binary terms (Eq. ~\ref{eq:unary_t}, ~\ref{eq:binary_t}) across every frame. We increase the number of replacement parts until the maximum allowed error value is reached. As seen in the figure, the number of replacement parts increases rapidly (from 2 to 100) for 5000 frames or less. However, an additional 5000 frames only leads to a small increase in dictionary size (from 100 to 115), affirming that a reasonably small number of replacement heads can capture the entire expressive range of a character. 

\paragraph{Limitations}
Our system is the first to address the technical problems of a stop-motion animation workflow and has limitations, subject to future work:
\begin{itemize}[leftmargin=*]
\item Our segmentation approach does not respect the aesthetics of a \singlepart boundary. While our approach seamlessly connects different parts, the deformation albeit minimal, can adversely impact geometric detail near the part boundary.
\item Despite a seamless digital connection, seams between physically printed \parts remain visible. Commerical animations often remove these seams digitally by image processing, but film directors like Charlie Kaufman have also made these seams part of the character aesthetic~\cite{Murphy:2015}. 
\item Physically re-assembling the object for every frame 
sequentially from a large replacement \library of \pieces can still be cumbersome. This could be addressed by a scheduling algorithm that proposes an order in which to shoot animation frames that minimizes object re-assembly. 
\item Our replacement part algorithm results are based on vertex position or deformation space distance metrics. We believe our results could be better using a perceptually based distance metric between instances of a deformable object.
\item \rinatchange{Currently our segmentation algorithm does not explicitly enforce symmetry.
Symmetry may sometimes be a desirable property that could be incorporated.
However, breaking symmetry has its advantages: the tuft of hair on the Camel's head in Fig. \ref{fig:faces_multi} is assigned to a single --- albeit symmetry-breaking --- \singlepart.}
\end{itemize}

%
%

\section{Conclusion}
Stop-motion animation is a traditional art-form that has seen a surge of
popularity with the advent of 3D printing.  Our system is the first attempt at an
end-to-end solution to the research problems in the creation of stop-motion
animations using computer animation and 3D printing. We hope this paper will
stimulate new research on the many problems encountered in the area of
stop-motion animation.

\section{Acknowledgements}
This work is funded in part by NSERC Discovery, the Canada Research Chairs
Program, Fields Institute CQAM Labs, and gifts from Autodesk, Adobe, and MESH.
We thank members of the dgp at University of Toronto for discussions and draft
reviews.

\bibliographystyle{ACM-Reference-Format}
\bibliography{stopmotion-acmtog}

\end{document}